# Metallic water: transient state under ultrafast electronic excitation

Nikita Medvedev[1,2,*], Roman Voronkov[3], Alexander E. Volkov[3,4]

[1] *Department of Radiation and Chemical Physics, Institute of Physics, Czech Academy of Sciences, Na Slovance 1999/2, 182 21 Prague 8, Czech Republic*

[2] *Laser Plasma Department, Institute of Plasma Physics, Czech Academy of Sciences, Za Slovankou 3, 182 00 Prague 8, Czech Republic*

[3] *P.N. Lebedev Physical Institute of the Russian Academy of Sciences, Leninskij pr., 53,119991 Moscow, Russia*

[4] *National Research Centre „Kurchatov Institute", Kurchatov Sq.1, 123182 Moscow, Russia*

## Abstract

The modern means of controlled irradiation by femtosecond lasers or swift heavy ion beams can transiently produce such energy densities in samples that reach collective electronic excitation levels of the warm dense matter state where the potential energy of interaction of the particles is comparable to their kinetic energies (temperatures of a few eV). Such massive electronic excitation severely alters the interatomic potentials, producing unusual nonequilibrium states of matter and different chemistry. We employ density functional theory and tight binding molecular dynamics formalisms to study the response of bulk water to ultrafast excitation of its electrons. After a certain threshold electronic temperature, the water becomes electronically conducting *via* the collapse of its band gap. At high doses, it is accompanied by nonthermal acceleration of ions to a temperature of a few thousand Kelvins within sub-100 fs timescales. We identify the interplay of this nonthermal mechanism with the electron-ion coupling, enhancing the electron-to-ions energy transfer. Various chemically active fragments are formed from the disintegrating water molecules, depending on the deposited dose.

## 1. Introduction

Being omnipresent in biological materials, water is the prototypical sample for studies of living organisms. The response of water to various types of radiation is used as the standard in radiation safety research and radiation therapy [1–5]. The disintegration of water under irradiation into various fragments (different species) – water radiolysis – was intensively studied over decades [1–5]. It plays a crucial role in biochemistry, defining survival rates of bio-molecules embedded in water solutions – in particular, living cells. Response of water molecules to high-energy photons such as X-rays and gamma rays, electrons, and fast ions (in material science called swift heavy ions, SHI, also known in bioscience as HZE, which stands for high-Z and energy) was a subject of numerous research [2,6–8].

Typically, it is assumed that each incoming high-energy particle produces a set of well-defined fragments embedded in an unaffected water environment. A small number of water molecules disintegrate into typical products: free radicals H, OH, O, and free electrons, and further chemical kinetics may lead to damage to other water molecules and create secondary products via interaction of the primary ones (such as $H_2O_2$, $H_2$, $O_2$, etc.) [9]. Since different radicals damage biological materials at different rates, it is important to trace the evolution of concentrations of them: e.g., the most damaging for human cells are considered reactive oxygen species (ROS) such as hydroxyl ($OH^{\cdot-}$) and superoxide ($O_2^{\cdot-}$) [10–12].

---

[*] Corresponding author: ORCID: 0000-0003-0491-1090; Email: nikita.medvedev@fzu.cz

The processes of formation and evolution of various kinds of radicals are usually described with a set of chemical rate equations, or corresponding kinetic Monte Carlo procedures [9,13]. Either method is based on the assumption that the chemical kinetics of created species takes place in normal water media. These conditions presume low fluxes (dose rates) of incoming radiation, such that the bulk of the water is either unaffected or has sufficient time to relax between impacts of successive particles.

These conditions are not satisfied in many irradiation scenarios. Modern femtosecond lasers, especially X-ray free-electron lasers (XFEL) produce extremely short and intense pulses, which ionize the target to high degrees, such that the densities of excited electrons approach the densities of atoms [14,15]. In SHI impacts, in the nanometric proximity of the ion trajectories, comparable levels of excitation may transiently be produced [16,17].

Excitation of a massive amount of electrons leads to the formation of unusual phases in materials [18,19]. When electrons are excited to high temperatures, whereas ions of the target transiently stay relatively cold, the matter departs from the equilibrium phase diagram and forms new states, unachievable otherwise. The physical and chemical characteristics of such states may be drastically different from the equilibrium phases and require dedicated studies. Recently, it has been reported that water irradiated with a femtosecond XFEL pulse becomes highly electronically conducting [20]. It was attributed to the transition of water into a liquid metal state, however, the exact mechanisms were unclear since the results of DFT simulations did not agree with the experiment [20].

Here, we demonstrate that under ultrafast excitation of the electronic system, bulk water transiently forms a metallic state *via* the collapse of the band gap, making it electronically conducting. This transition is induced by an interplay of the nonthermal modification of the interatomic potential and enhanced electron-ion coupling. It is accompanied by the softening of the interatomic potential, eventually leading to the disintegration of the water molecules into various species, depending on the deposited dose (electronic temperature reached): at low doses, the common reactive species such as OH are produced, whereas, with the increase of the dose, they fragment into atomic H and O species.

## 2. Models

Upon ultrafast irradiation, such as with XFEL or SHI, primarily the energy is deposited into the electronic system of the target, whereas its ionic system stays cold [14,21,22]. The material response started with the excitation of electrons, and proceeds *via* secondary cascades, exciting new electrons from the valence band or core shells to the conduction band of the material. At the same time, created core-shell holes decay via the Auger channel, also exciting secondary electrons [23]. Secondary electron cascades typically finish within a few (or a few tens) of femtoseconds, when all electrons lose their energy below a certain threshold and cannot ionize new electrons [24,25]. The majority of electrons are in local thermodynamic equilibrium even during the cascading stage, with only a minority of electrons performing the cascades [26,27].

With femtosecond pulse duration and fluences typically produced at XFELs, the electronic system of the target can reach temperatures of a few electron-Volts [14,15,28]. At such densities and energies of the electronic ensemble, it quickly thermalizes and forms the two-temperature state: with an electronic temperature high above the atomic one [14,29].

The effect of the high-temperature electronic system is twofold: it affects the interatomic potential - collective potential energy surface in the material (nonthermal effect), and it starts to exchange energy with ions *via* electron-ion coupling. The first one may lead to molecular destabilization and result in nonthermal phase transitions, for example, nonthermal melting [30,31]. These effects can be studied with *ab-initio* methods, such as density-functional theory molecular dynamics (DFT-MD) within the Born-Oppenheimer (BO) approximation [32]. It assumes that electrons are much lighter and faster than the ions and thus electronic wave functions instantly adjust to atomic displacements. The BO approximation, thus, does not allow for non-adiabatic coupling between electronic and atomic movements (electron-ion scattering, including normal atomic oscillations) which triggers electronic transitions between their energy levels, thereby exchanging energy between the electrons and ions [33]. The latter process is responsible for the equilibration of the electronic and atomic temperatures in the two-temperature state [14].

Here, we employ two different methods to study the abovementioned effects. The first *ab-initio* method is the DFT-MD, which relies on the BO approximation. It allows us to study precisely the effects of electronic excitation to high temperatures on the interatomic forces and the response of the ionic system to it. The second one is based on tight-binding molecular dynamics (TBMD). It is a more approximate method than DFT, but it allows us to independently validate our conclusions, treat larger systems, and, most importantly, study the nonadiabatic coupling effects. Below we briefly describe the two methods.

Both approaches enable calculations of the evolution of the electronic energy levels (molecular orbitals, or band structure), the electronic wave functions, and the potential energy surface - forces acting on atoms. Excitation of electrons directly affects all these, thereby allowing us to take into account nonthermal effects.

The DFT-MD tool is employed *via* the Quantum Espresso simulation package [34]. Within this approach, Kohn-Sham one-particle equations are solved at each MD step to obtain electronic energy levels and to compute interatomic forces [35]. Perdew-Burk-Ernzerhof's (PBE [36]) approximation for the exchange-correlation energy functional – the key parameter in DFT – was chosen to simulate a 96-atom cell. Although it underestimates band gaps at ambient conditions, it performs much better at elevated electronic temperatures [37] and provides qualitatively the same results as more accurate but slower-performing meta-GGA functionals [38].

To remove core electrons from the simulation we use fhi98PP norm-conserving pseudopotentials [39] from the ABINIT website library [40]. The size of the plane wave basis set is controlled by energy cutoff $E_{cutoff}$ = 170 Ry (2313 eV). The molecular dynamics use the Verlet algorithm with the time step 0.5 fs.

The second tool used here is the earlier developed hybrid code XTANT-3 [14,41]. In XTANT-3, the electronic band structure is calculated using the transferrable tight binding (TB) method [14]. We applied mio-1-1 DFTB parameterization which uses an $sp^3$ basis set for the description of water, specifically developed to describe biological systems [42].

To trace the nonadiabatic energy exchange between the electrons of the valence and conduction band with atoms, XTANT-3 uses Boltzmann collision integrals (BCI) formalism [41]. The electron-ion collision integral depends on the transient matrix elements of the electron-ion interaction, and the electron populations (distribution function).

The atoms are traced with the molecular dynamics (MD) simulation. The interatomic potential, evolving together with the electronic system, is provided by the transient TB Hamiltonian. The energy transferred from electrons calculated with the BCI is fed to atoms *via* the velocity scaling algorithm at each time step of the simulation [41]. We apply the $4^{th}$-order Matryna-Tuckerman algorithm to propagate atomic coordinates with the timestep of 0.05 fs [43]. Typically, 192 atoms are used in the simulation box with periodic boundary conditions.

For both DFT-MD and TBMD simulations, before productive runs, the water state was created with the following procedures: water molecules were placed randomly in the simulation box and adjusted in size to produce normal water density for the chosen number of molecules. The starting rotational orientation of molecules was also chosen randomly. Then, the system was relaxed with the steepest descent algorithm to reach the minimum of the potential energy. After that, the simulations were initialized with atoms in those positions and random velocities assigned according to the Maxwellian distribution (at room temperatures). Before the increase of the electronic temperature, mimicking the arrival of an X-ray pulse (with the chosen duration of 10 fs FWHM), the system was allowed to equilibrate for a few hundred femtoseconds [41].

XTANT-3 has previously been validated by comparisons with experimental data for various materials, see e.g. Refs. [14,41,44,45], and thus we assume that it can be applied here to water. We validate its performance at elevated electronic temperatures against the DFT-MD calculations in the next section.

## 3. Results
### 3.1 Born-Oppenheimer simulation: nonthermal disintegration

DFT-MD simulations predict changes in water band structure starting from $T_e$~2.5 eV (1.6 eV/atom, 5.1% of valence electrons excited to the conduction band). At this dose (see Figure 1a), a part of the conduction band energy levels starts to descend at ~380 fs forming impurity-like levels in the band gap.

Further increase of the electronic temperature results in faster creation of such levels also making the energy spectrum inside of the former bandgap denser. At $T_e$~2.75 eV (2 eV/atom, 6.2% of valence electrons excited to the conduction band) first levels of the conduction band reach the top of the valence band as fast as ~50 fs (Figure 1b).

At last, at $T_e$~3 eV (2.5 eV/atom, 7.2% of valence electrons excited to the conduction band) the band gap collapses completely within ~25 fs merging the valence and the conduction bands into a quasi-continuous energy spectrum (Figure 1c). This indicates a formation of the electronically conducted state, a liquid metal.

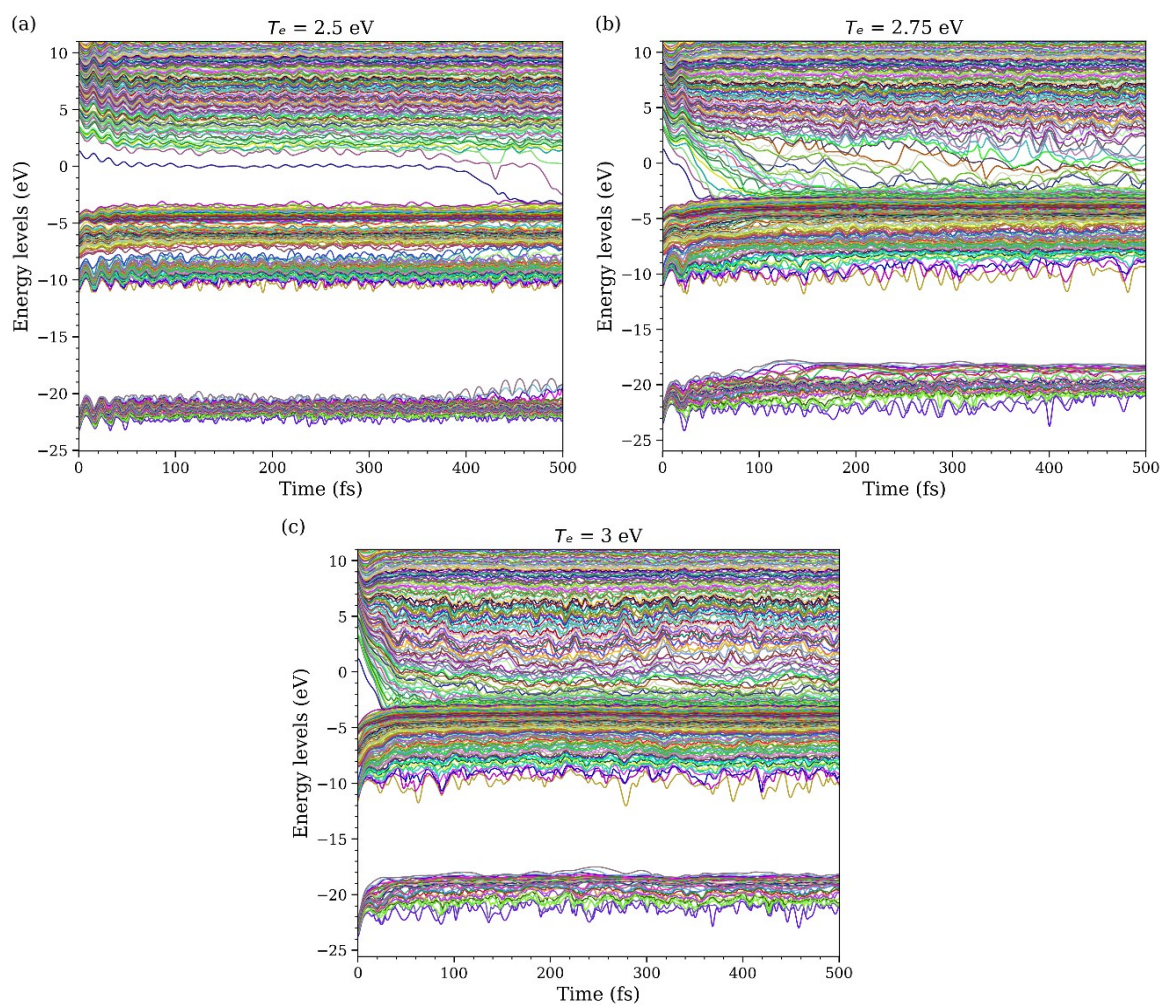

Figure 1. Electronic energy levels evolution of water under ultrafast energy deposition with the dose of (a) 1.6 eV/atom, (b) 2 eV/atom, and (c) 2.5 eV/atom calculated with Quantum Espresso code.

Figure 2 demonstrates the radial pair distribution functions of water at $T_e$~3 eV. A significant decrease of the O-H correlation peak at ~1 Å indicates that the vast majority of O-H bonds are broken already at ~25 fs after electronic temperature elevation which coincides with the timescale of the band gap collapse. This indicates that the band gap stability depends on the stability of molecular bonds rather than on the mutual positioning of water molecules.

The rise of the H-H correlation peak (Figure 2b) at ~1 Å may indicate attempts to form $H_2$ molecules by detached hydrogens, while the O-O function remains almost intact indicating that oxygen atoms freed from bonds with hydrogen are not trying to recombine into $O_2$ molecules at these timescales.

These conclusions are confirmed by an analysis of fragments appearing in the simulation box during the simulation. The fragments are identified by an analysis of neighboring atoms within cutoff distances 1.15 Å for H-O, 1.8 Å for O-O and 0.9 Å for H-H bonds (changing these values only trivially alter the fragmentation picture). As it is demonstrated in Figure 3, the main constituents of the box are atomic hydrogen and oxygen with the moderate presence of $OH^-$ and $H_2$ molecules and the occasional transient appearance of $H_3$ clusters. The short-lived appearance of such species indicates a liquid flow of hydrogens, which may transiently approach each other forming hydrogen clusters and then drifting apart.

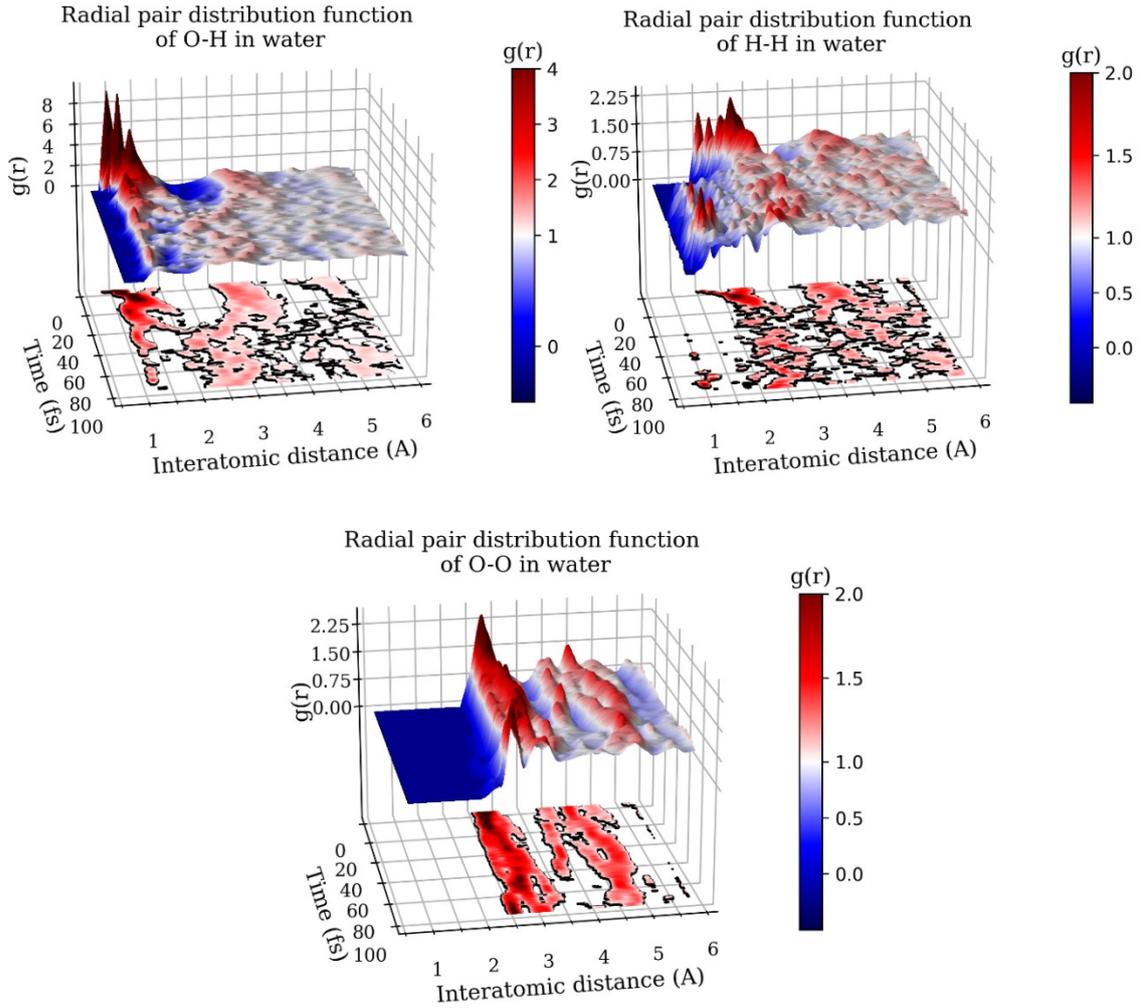

*Figure 2. Radial pair distribution function of (a) O-H, (b) H-H, and (c) O-O atomic pairs in water under ultrafast energy deposition with the dose of 2.5 eV/atom ($T_e$~3 eV) calculated with Quantum Espresso code. Panels under surfaces demonstrate the same information in a form of a colormap with g(r) values below 1 removed to highlight correlation peaks.*

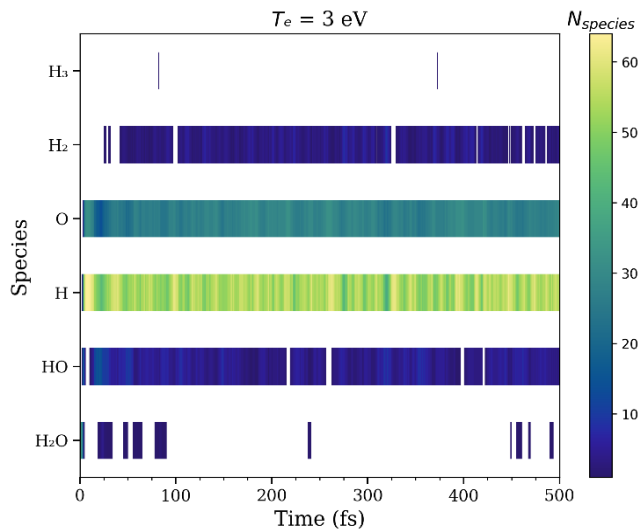

*Figure 3. Number of various fragments in the simulation box under ultrafast energy deposition with the dose of 2.5 eV/atom ($T_e$~3 eV) calculated with Quantum Espresso code.*

The potential energy released during the band gap collapse converts to the kinetic energy of the atomic system [46]. Due to this nonthermal acceleration of atoms, the atomic temperature (kinetic temperature) may reach far above the water evaporation point, see Figure 4. Due to BO approximation, no electron-ion kinetic energy exchange is present, and instantaneous atomic temperature stabilizes at ~1900 K at electronic temperature $T_e$~3 eV.

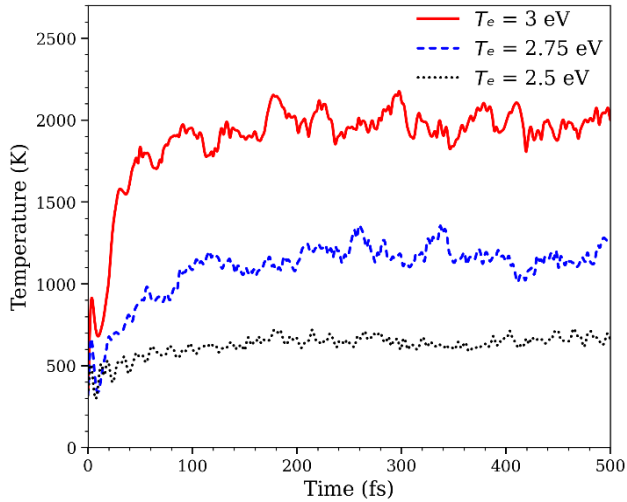

*Figure 4. Instantaneous atomic kinetic temperatures at different electronic temperatures calculated with Quantum Espresso code.*

The quality of the DFTB parameterization mio-1-1 for water under normal conditions was extensively studied in the literature [47,48]. Here, we evaluate its performance at high electronic temperatures. The calculated threshold of the band gap collapse within the BO approximation is $T_e$~3.7 eV (corresponding to 6.1 eV/atom deposited dose, or ~15% of valence electrons excited to the conduction band). Similar to the described above DFT results, the collapse of the band gap starts with the shrinkage of the entire gap and the appearance of the first levels inside of it, induced by a detachment of hydrogens from the water molecules. As the interatomic potential binding hydrogen and oxygen atoms become shallower with an increase of the electronic temperature, at the threshold dose it becomes barely binding. Collisions of the water molecules in the liquid phase at room atomic temperature knock out the first "seed" hydrogen atom, which then knocks out the next one, forming new defect energy levels inside the gap.

With the increase of the dose (electronic temperature), more and more water molecules disintegrate, and eventually, the entire band gap disappears, forming merged valence and conduction band. Thus, in TBMD simulation, the state also becomes metallic (electronically conducting). The XTANT-3 calculations qualitatively agree with the DFT-MD calculations, even though quantitatively the damage threshold is higher. We thus conclude that the DFTB method can be applied for the qualitative analysis of water at high electronic temperatures. In the next section, we employ it beyond the BO approximation to study nonadiabatic effects and their interplay with the nonthermal damage reported here.

### 3.2 Nonadiabatic effects: synergistic heating

Modeling water response accounting for the nonadiabatic energy exchange between electrons and ions (electron-ion coupling, e.g. scattering) in water noticeably reduces the damage threshold down to the dose of ~1.1 eV/atom (peak electronic $T_e$~1.5 eV, peak density of excited electrons ~2.8 %), as calculated with XTANT-3. The damage takes place *via* an interplay of the nonthermal and nonadiabatic effects, similar to the reported one in solid oxides and polymers [18,49]. The initial kick to atoms due to nonthermal modification of the interatomic potential leads to a very fast increase in the atomic temperature (as described in the previous section). An increased atomic temperature leads to an increase in the electron-ion coupling, triggering even faster energy exchange [46]. This self-amplifying process increases the atomic temperature at ultrashort, sub-picosecond, timescales, see the example in Figure 5. The average atomic temperatures reach ~5000 K after equilibration with the electronic one.

It is interesting to note that the oxygen temperature transiently increases higher than the hydrogen one, indicating that the oxygens experience stronger modifications of the interatomic forces due to electronic excitation to high temperatures. However, the deviations quickly disappear, and within a few hundred femtoseconds, the two species are equilibrated again.

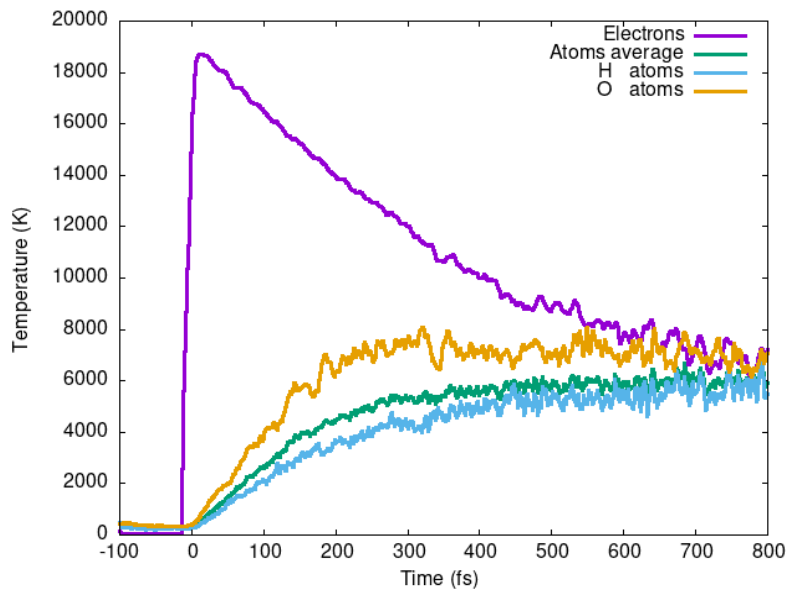

*Figure 5. Electronic and atomic temperatures (average and partial of hydrogen and oxygen systems) of water under ultrafast energy deposition with the dose of 1.1 eV/atom (10 fs FWHM centered at t=0 fs), performed with XTANT-3 code.*

The extremely fast heating leads to molecular fragmentation, which is accompanied by the formation of the defect levels within the bandgap, similar to the BO-case in the last section. Here, the bandgap completely collapses at the deposited dose of ~1.5-1.7 eV/atom (peak electronic temperatures of $T_e$~20.000K=1.7 eV), see Figure 6. Water, thus, turns into a metallic liquid state at lower doses than those produced within BO approximation.

Note that this behavior and the electronic temperature coincide very well with the experimental report of the metallic water formation in Ref. [20]. As was also noted in Ref. [20], the simulations within BO approximation do not produce conducting state at the experimental conditions ($T_e$~20000 K), which agrees with our estimate in the last section. We thus conclude that the liquid metallic water is produced *via* an interplay of the nonthermal and nonadiabatic effects, which require simulations beyond the BO approximation.

This process is accompanied by water fragmentation into various species, as shown in Figure 7. The majority of the produced fragments are hydrogens (H, $H_2$, with little contribution of $H_3$), and O and OH radicals, with some contribution of $O_2$ molecules to which floating hydrogen may attach and detach. With the increase of the dose (electronic temperature) above ~ 1.5 eV/atom, the concentration of the OH radical reaches a plateau and new OH radicals stop forming, instead fragmenting into H and O (see Appendix).

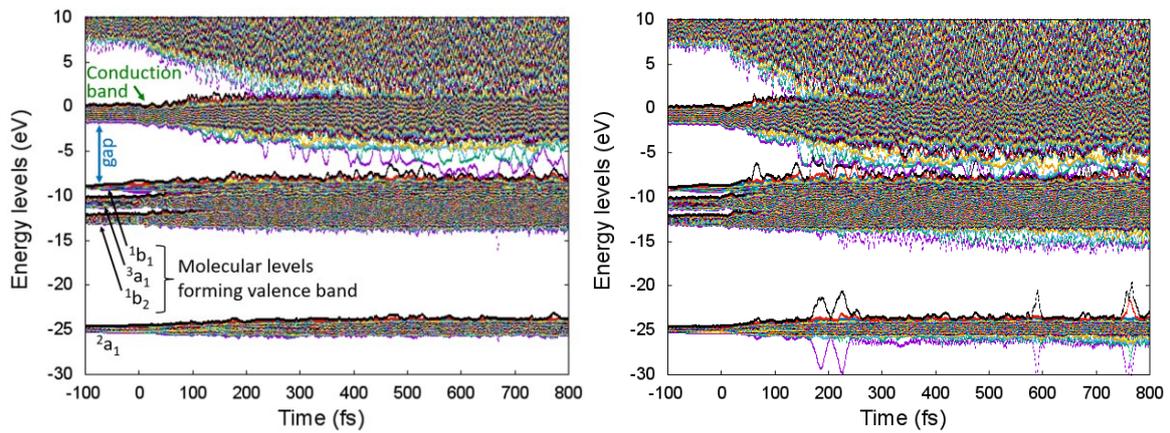

*Figure 6. Electronic energy levels evolution of water under ultrafast energy deposition with the dose of 1.1 eV/atom (left panel) and 1.7 eV/atom (right panel), performed with 10 fs FWHM pulse centered at t=0 fs, calculated with XTANT-3 code.*

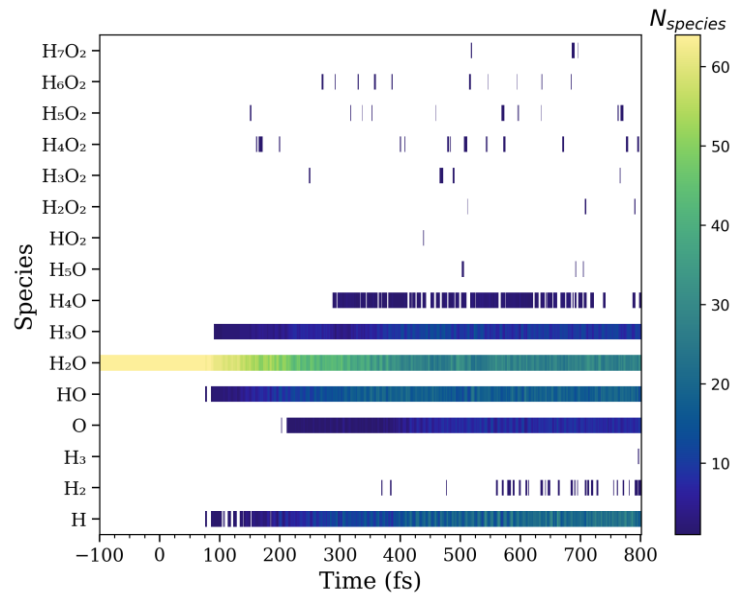

*Figure 7. Relative populations of various species in the simulation box of water under ultrafast energy deposition with the dose of 1.1 eV/atom, performed with 10 fs FWHM pulse centered at t=0 fs, calculated with XTANT-3 code.*

## 4. Discussion

Having evaluated the damage threshold for the formation of metallic water, we now proceed with an estimate of the achieved doses after an SHI impact. For that, we apply the TREKIS-3 Monte Carlo code, which evaluates the ion energy loss in a given matter, secondary electron transport induced, Auger and radiative decays of core holes produces, transport of valence holes and emitted photons [50,51]. We estimate the energy density in the ensemble of the excited electrons, and the energy transferred to ions after the impact of the U ion with energy of 2000 MeV, see Figure 8. Following our previous works [46,52], energy transfer to ions is assumed to take place via three channels: elastic scattering of excited electrons, elastic scattering of valence holes, and release of the potential energy of electron-hole pairs associated with the nonthermal acceleration described in Section 3.1 above. This way, both nonthermal and nonadiabatic energy transfers may be approximately extracted from a Monte Carlo simulation, as discussed in Refs. [46,53].

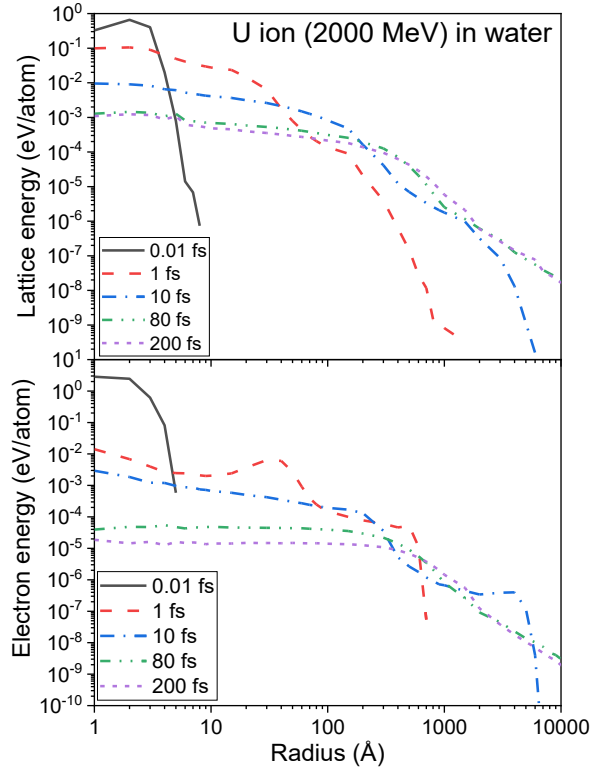

*Figure 8. Energy transferred to atoms via elastic scattering of excited electrons and valence holes with the added energy of electron-hole pairs (top panel), and kinetic energy of excited electrons (bottom panel) in water at different times after the impact of the U ion (2000 MeV), simulated with TREKIS-3 [50,51].*

The results in Figure 8 suggest that even after the heaviest ion impact with the energy near its Bragg peak (maximal energy loss, $S_e$~16 keV/nm), the energy densities reached in the electronic and ionic system in water are insufficient to produce a metallic state. We thus conclude that the effects of impacts of swift heavy ions in water may be described within the standard models, assuming the production of radicals in a regular insulating water environment [9].

## 5. Conclusion and outlook

We studied water under ultrafast deposition of energy into its electronic system using two methods: density functional and tight binding molecular dynamics. Both methods applied within the Born-Oppenheimer (BO) approximation predicted the formation of metallic liquid water above a certain dose (electronic temperature ~3-4 eV). Including nonadiabatic (non-BO) energy exchange between electrons and ions lowers the threshold dose to ~1.1 eV/atom. In this case, the collapse of the band gap and formation of the metallic state is induced by an interplay of the nonthermal and nonadiabatic effects: acceleration (heating) of atoms due to modification of the interatomic potential by the electronic excitation enhances electron-ion coupling. Reaching high ionic temperatures at sub-picosecond timescales, together with high electronic temperatures, also leads to fragmentation of water molecules into various fragments. We show that the formation of OH radicals reaches a plateau at the doses of ~ 1.5 eV/atom, above which mostly H and O atomic species are formed. Our results on ultrafast formation of the metallic liquid state in water at electronic temperatures ~20000 K are supported by recent experiments [20].

## 6. Conflict of Interest

The authors declare no conflict of interest, financial or otherwise.

## 7. Data Availability Statement

The data that support the findings of this study are available from the corresponding author upon reasonable request.

## 8. Acknowledgments

NM gratefully acknowledges financial support from the Czech Ministry of Education, Youth and Sports (grant No. LM2018114). RV and AEV were funded by a grant from Russian Science Foundation No. 22-22-00676. Computational resources for XTANT-3 calculations were supplied by the project "e-Infrastruktura CZ" (e-INFRA LM2018140) provided within the program Projects of Large Research, Development and Innovations Infrastructures. DFT calculations have been carried out using computing resources of the federal collective usage center Complex for Simulation and Data Processing for Mega-science Facilities at NRC "Kurchatov Institute", http://ckp.nrcki.ru/.

## 9. Appendix

The electronic energy levels in water after energy deposition, corresponding to the electronic temperatures of $T_e$~3.7 eV and 4 eV, calculated with TBMD (XTANT-3) in BO approximation are shown in Figure 9. A qualitative comparison with DFT-MD calculations, shown in Section 3.1 (cf. Figure 1), demonstrates that the mio-1-1 DFTB approximation can be used to describe water at high electronic temperatures. However, overestimation by the absolute value should be kept in mind for comparisons with experiments.

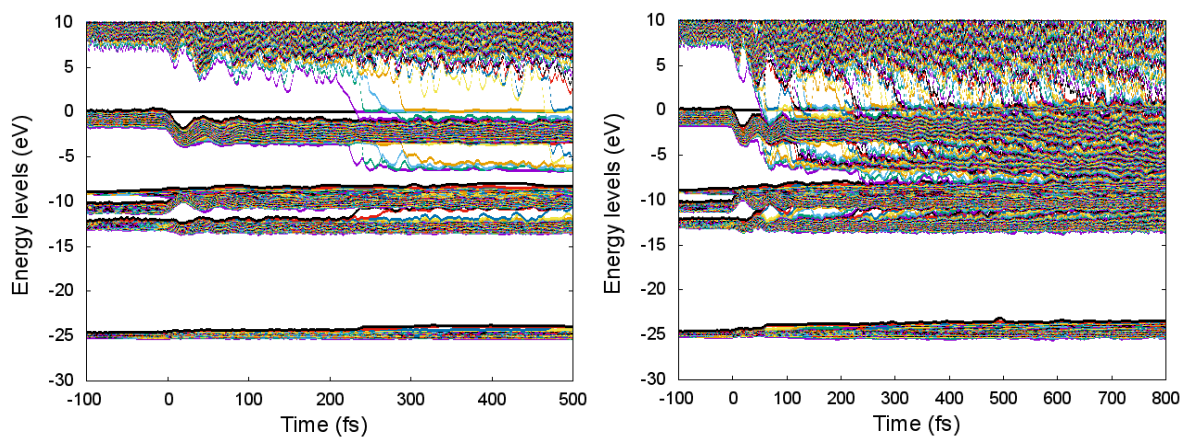

*Figure 9. Electronic energy levels in water after energy deposition of 6.1 eV/atom ($T_e$~3.7 eV; left panel), and 6.5 eV/atom ($T_e$~4 eV; right panel), calculated with TBMD (XTNAT-3) in BO approximation.*

The relative concentrations of various atomic and molecular species in water after irradiation with doses from 0.5 eV/atom to 2 eV/atom are shown in Figure 10. The concentration of intact water molecules decreases with the dose increase. The fraction of OH radicals first increases with the increase of the deposited dose, then reaches its maximum at the dose of ~1.5 eV/atom. At high doses, OH radicals become unstable and dissociate into atomic H and O. Other molecular fragments almost do not form at the timescales studied here.

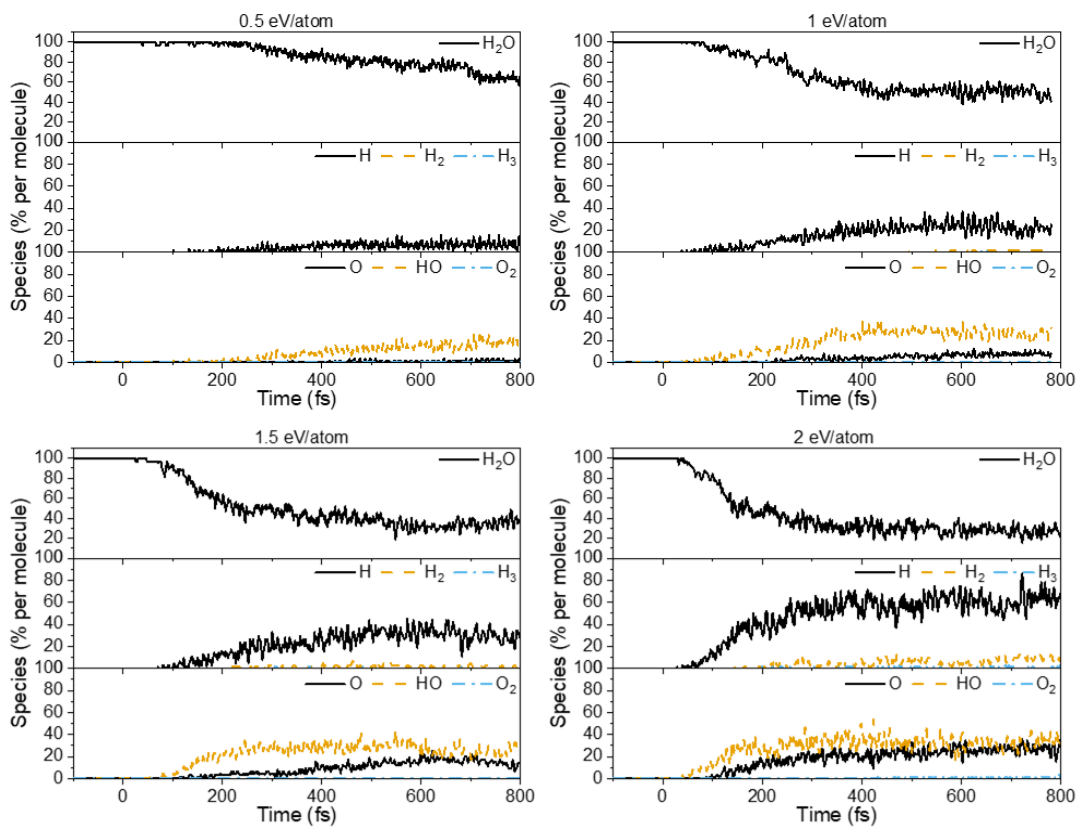

*Figure 10. Contribution of various species in water after irradiation with doses of 0.5 eV/atom (top left panel), 1 eV/atom (top right panel), 1.5 eV/atom (bottom left panel), 2 eV/atom (bottom right panel), modeled with XTANT-3 code.*